\documentclass[a4paper,onecolumn,11pt,accepted=2024-02-01]{quantumarticle}
\pdfoutput=1
\usepackage[utf8]{inputenc}
\usepackage[english]{babel}
\usepackage[T1]{fontenc}
\usepackage{amsmath}
\usepackage{hyperref}
\usepackage{amssymb}
\usepackage{tikz}
\usepackage{lipsum}
\usepackage{xspace} 
\usepackage{float}

\newcommand{\Tr}{\mathrm{Tr}}

\definecolor{darkgreen}{rgb}{0.0, 0.2, 0.13}

\def\({\left (}
\def\){\right )}
\def\tr{\text{tr}}

\usepackage[numbers,sort&compress]{natbib}
\newcommand\mydots{\hbox to 1em{.\hss.\hss.}}

\graphicspath{{Figure/PNG/}{Figure/PDF/}{Figure/EPS/}{Figure/TEX/}{Figure/}}

\begin{document}

\title{Microcanonical windows on quantum operators}

\author{Silvia Pappalardi}
\orcid{0000-0002-2445-2701}
\email{pappalardi@thp.uni-koeln.de}
\affiliation{Laboratoire de Physique de l’\'Ecole Normale Sup\'erieure, ENS, Universit\'e PSL, CNRS, Sorbonne Universit\'e, Universit\'e de Paris, F-75005 Paris, France}
\affiliation{Institut f\"ur Theoretische Physik, Universit\"at zu K\"oln, Z\"ulpicher Straße 77, 50937 K\"oln, Germany}

\author{Laura Foini}
\affiliation{IPhT, CNRS, CEA, Universit\'e Paris Saclay, 91191 Gif-sur-Yvette, France}

\author{Jorge  Kurchan }
\affiliation{Laboratoire de Physique de l’\'Ecole Normale Sup\'erieure, ENS, Universit\'e PSL, CNRS, Sorbonne Universit\'e, Universit\'e de Paris, F-75005 Paris, France}

\maketitle

\begin{abstract}We discuss a construction of a microcanonical projection WOW of a quantum operator O induced by an energy window filter W, its spectrum,
and the retrieval of canonical many-time correlations from it.
\end{abstract}

\tableofcontents

\section{Introduction}
\begin{figure*}[t]
	\centering
    \includegraphics[width=1\linewidth]{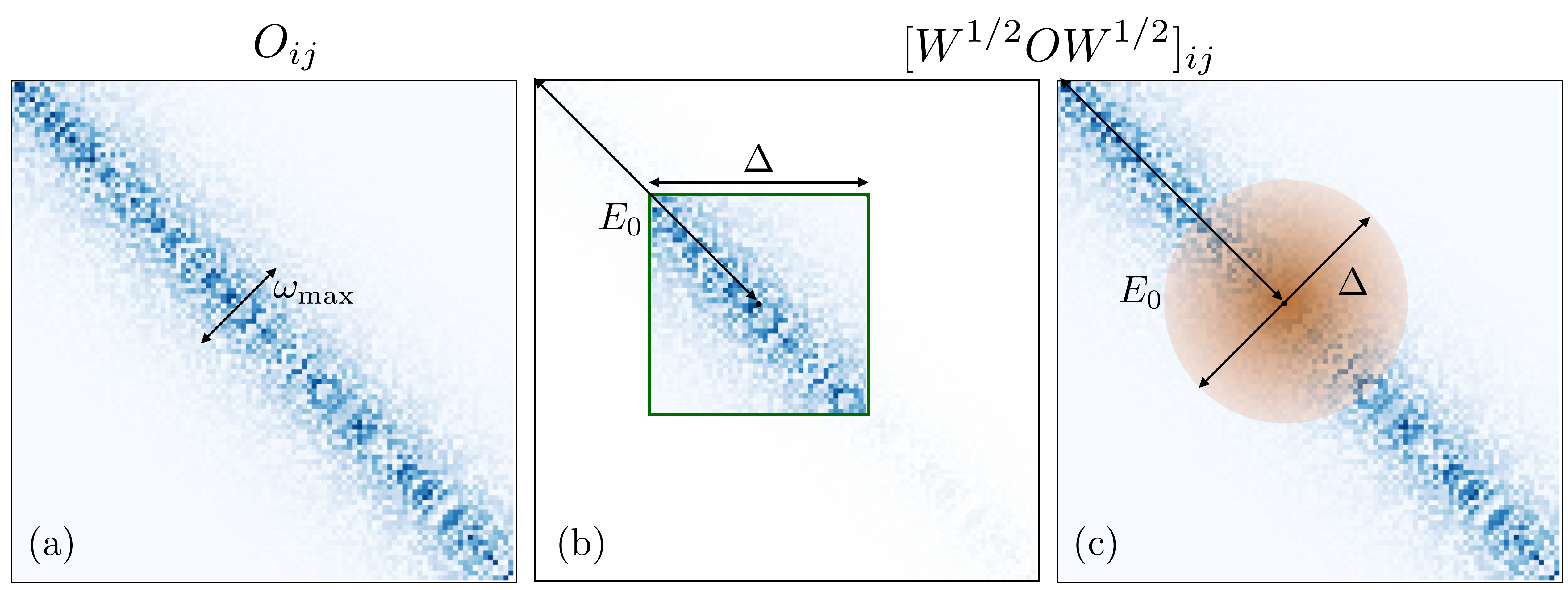}    
	\caption{Pictorial representation of the observable $O$ and of the microcanonical projected  WOW. (a) The matrix elements in the energy eigenbasis $O_{ij}=\langle E_i|O|E_j\rangle$ look like structured random matrices characterized by an intrinsic, observable dependent, energy scale $\omega_{\max}$ after which they decay to zero.
 (b-c) Observable $W^{1/2}O W^{1/2}$ projected in an energy window around $E_0$ with width $\Delta$. 
 (b) Box filter $W$: the microcanonical operator is given only matrix elements with energies inside the box $|E_{i,j}-E_0|<\Delta/2$. (c) Gaussian filter $W$: the observable is multiplied by a Gaussian function in both directions $i,j$. }
	\label{fig:WOW}
\end{figure*}

 Quantum thermalization and dynamics at equilibrium are nowadays well understood in terms of the matrix elements of physical operators in the energy eigenbasis, i.e. $O_{ij}=\langle E_i|O|E_j\rangle$ \cite{deutsch1991quantum, srednicki1999approach, dalessio2016from}. These behave as  random matrices with correlated elements, whose smooth statistical properties encode all the physical features \cite{foini2019eigenstate}.  Understanding the nature of the randomness in $O_{ij}$ and how to extract from them or their spectrum meaningful information is still an open, largely debated issue \cite{fyodorov1991scaling,kus1991density,fyodorov1996wigner,prosen1994statistical, cotler2017chaos, dymarsky2019new, dymarsky2019mechanism, dymarsky2022bound, richter2020eigenstate, wang2022eigenstate, brenes2021out}. 
The purpose of this paper is to discuss the construction of a microcanonical projection
of an operator $O$, in the thermodynamic limit: 
\begin{equation}
\label{WOW}
    \tilde O \equiv  W^{1/2}  O \,  W^{1/2} \ ,
\end{equation}
referred to as WOW, where $W$ is some `window' operator which is centred around an energy $E_0$, as illustrated in Fig.\ref{fig:WOW}.

The rules we set for this exercise are that {\it (i)} the  projected operator $\tilde O$
is such that its dynamical correlators among $n$-times allow one to retrieve the corresponding canonical
$n$-time correlators of $O$, and that {\it (ii)} the spectrum of $\tilde O$
provides a meaningful large-deviation function.
For example, defining the commutator operator
$${\cal{C}} \equiv [\tilde O(t), \tilde O]_-$$
the spectrum of the operator ${\cal{C}}$ contains all the `multifractality' properties of the quantum generalized Lyapunov exponent, through its moments $\Tr \left\{ {\cal{C}}^{2n} \right\}$, for $n>0$ \cite{pappalardi2023quantum}.\\

It turns out that for this to work, as is to be expected, the width of the energy window is not arbitrary, but, somewhat surprisingly, also its smoothness properties  have important consequences. 
We exemplify numerically our findings on the Ising model with transverse and longitudinal fields.

\section{Why WOW: a summary of the results}
 We set ourselves to construct an operator which is a microcanonical restriction of $O$ around some energy $E_0$.
This operator shall not only give the microcanonical expectation value, i.e. $\langle E_0|O|E_0\rangle$, but should also reproduce all correlations \emph{on the energy shell} of $E_0$.\\ 

\subsection{Correlations on the energy shell}
A way to introduce  correlations on an energy shell is to look at the following \emph{canonical regularized correlator}
\begin{equation}
    \label{regulabello}
    M_n(\vec t) = \Tr \left(\rho^{\frac 1n} O(t_1) \rho^{\frac 1n} O(t_2) \dots \rho^{\frac 1n} O(t_n)\right) \ ,
\end{equation}
where $\rho =e^{-\beta H}/Z$ and $Z = \text{Tr}(e^{-\beta H})$ with $\beta$ the inverse temperature, and $O(t) = e^{-i H t} O e^{-i H t}$ for any (eventually complex) $\vec t = (t_1, t_2, \dots t_n)$ times \footnote{We have absorbed here $\hbar$ into the $\vec t$, so that time has units of inverse energy. }, where we set $\hbar=1$. While familiar in high energy physics \cite{maldacena2016bound}, these regulated correlation functions may seem 
odd for a statistical physicist. To mitigate the understandable prejudice, let us note that
this correlator is related to the more usual thermal average
\begin{equation}
    \label{canonical}
    S_n(\vec t) = \Tr \left(\rho O(t_1)O(t_2) \dots O(t_{n-1}) O(t_n)\right)  \ ,
\end{equation}
by a simple shift in imaginary time: 
Defining $\vec \ell_n = (\frac{n-1}{n}, \frac{n-2}{n}, \dots, \frac{1}{n}, 0)$, one has 
\begin{equation}
    S_n(\vec t) = M_n(\vec t +i \beta \vec \ell_n) \ ,
\end{equation}
which in frequency $\vec \omega = (\omega_1, \dots, \omega_n)$ reads
\begin{equation}
   S_n(\vec \omega) =  M_n(\vec \omega) e^{-\beta \vec \ell_{n} \cdot \vec \omega}\ ,
\end{equation}
thus making explicit  the fluctuation-dissipation theorem \cite{haehl2017thermal}
\footnote{
With this shift, one rephrases the usual Kubo-Martin Schwinger (KMS) relation as a 
a time-reversal condition for Eq.\eqref{regulabello}: $M_n(t_1,...,t_n)=
\left\{M_n(t_n,...,t_1)  \right\}^*$.}.
 At small frequencies, the two correlators coincide. Thus, the regulated correlation function $M_n(\vec t)$ may be expected to share the same dynamics of the canonical one $S_n(\vec t)$ at large times, exceptions may be related to quantum bounds on time scales \cite{maldacena2016bound, tsuji2018bound, pappalardi2022quantum}. 

The \emph{thermal projection}
\begin{equation}
    \label{thermalproj}
    \rho^{ \frac 1{2n}} \, O(t) \, \rho^{\frac 1{2n}}
\end{equation} 
plays the role of constraining the support of each observable $O(t)$ on the thermal energy shell: the correlator in Eq.\eqref{regulabello} is evaluated on the shell identified by the thermal energy $E_\beta = \Tr(H e^{-\beta H})/Z$. Reproducing  this feature with a fixed, $n$-independent window is the focus of this article.\\

Another way to discuss on-shell corrections is via the \emph{quantum microcanonical ensemble}, nowadays well understood via the Eigenstate-Thermalization-Hypothesis (ETH) \cite{deutsch1991quantum, srednicki1999approach, dalessio2016from, foini2019eigenstate}.  The general form of ETH states that observables in the energy eigenbasis are pseudorandom matrices, whose statistical properties are smooth functions. 
Specifically, products of $n$ matrix elements with different indices read \cite{foini2019eigenstate}
\begin{equation}
    \label{ETHq}
    \overline{O_{i_1i_2}O_{i_2i_3}\dots O_{i_{n}i_1}} = e^{-(n-1)S(E^+)} F_{e^+}^{(n)}(\vec \omega) \ ,
\end{equation}
while products with repeated indices factorize in the large $N$ limit, in a manner described by free probability \cite{pappalardi2022eigenstate}. 
Here, $E^+=(E_{i_1}+\dots +E_{i_n})/n$ is the average energy, $\vec 
\omega = (\omega_{i_1i_2}, \dots, \omega_{i_{n-1}i_n})$ with $\omega_{ij}=E_i-E_j$ are $n-1$ energy differences. 
The $F^{(n)}_{e^+}(\vec \omega)$ are smooth functions of the energy density $e^+ = E^+/N$ and $\vec \omega$ and correspond to the microcanonical (on-shell) correlations of order $n$ at energy $e_+$. 

The relation between the microcanonical definition via Eq.\eqref{ETHq} and the canonical one in  Eq.\eqref{regulabello} is known. As reviewed below, the ETH smooth functions in Eq.\eqref{ETHq} can be expressed as the Fourier transform of the following correlator
\begin{align}
 \overline k_n(\vec t)  & \equiv
    \Tr' \left ( \rho^{\frac 1n} O(t_1) \rho^{\frac 1n} O(t_2) \dots \rho^{\frac 1n} O(t_n) \right ) 
    =    
    \int d\vec t e^{i \vec t \cdot \vec \omega}\,\,F^{(n)}_\beta (\vec \omega) 
    \label{kbarq}    
\end{align}
where the notation $ \Tr'$ indicates the trace constrained to matrix elements with different indices, e.g. $ \Tr'(A B C) = \sum_{i\neq j\neq k} A_{ij}B_{jk}C_{ki}$. These correlations are related to the thermal free cumulants \cite{pappalardi2022eigenstate} (a type of connected correlation function defined in Free Probability \cite{mingo2017free}) and constitute the building block of multi-time correlation functions. In Eq.\eqref{kbarq} and in what follows, we  use $\beta$ intended as an implicit function of the energy density $e_\beta=E_\beta/N$, e.g. $F^{(n)}_\beta (\vec \omega)= F^{(n)}_{e_\beta} (\vec \omega)$.

The behaviour of  all the $F_\beta^{(n)}(\vec \omega)$ as  functions of frequencies encode all the physical properties,
including some that are neglected if  elements are taken as independent \cite{foini2019eigenstate, pappalardi2022eigenstate}. In Hamiltonian systems, $F^{(n)}(\vec \omega)$ decays rapidly at large frequencies, reflecting the property that correlations between matrix elements with very different energies shall be small. 
The case $n=2$ has been related to the ``operator growth hypothesis'' which conjectures a universal exponential decay for $F^{(2)}_E(\omega)$ in the case of chaotic systems \cite{elsayed2014signatures, parker2019universal}, see also \cite{avdoshkin2020euclidean}. This structure has been discussed also for higher $n$ \cite{murthy2019bounds}:  at large frequencies, all $F^{(n)}(\vec \omega)$ should fall at least as
\[
F^{(n)}(\vec \omega) \sim e^{- |\omega_i|/\omega_{\max}}  \quad \text{in all directions } \omega_i \ ,
\]
where \cite{murthy2019bounds}
\begin{equation}
    \omega_{\max} \leq \frac 1{\beta \hbar} \ .
\end{equation}


\subsection{The problem}
The issue with the definition in Eq.\eqref{regulabello} is that the thermal projection \eqref{thermalproj} depends on $n$: each $n$-th correlation function needs a different projection. \\
This poses the question: can we replace instead, for the purposes of computing all $n$ on-shell correlation functions, 
the operator $O$ by a {\it single} projected version?

In this work, we consider an \emph{energy `window' operator} 
\begin{equation}
\label{W}
    W = W_\Delta(H - E_0)
\end{equation}
centered around energy $E_0$ with width $\Delta$, such that $[W,H]=0$.
The window is a filter in energy around $E_0$ and on the basis of the Hamiltonian reads
\begin{equation}
    \label{Wspec}
    W = \sum_i p_i |E_i\rangle\langle E_i| \ , \qquad p_i = p(E_i)\ge 0 \ ,
\end{equation}
with a distribution $p_i$, which we leave unnormalized. Standard choices include ``box projectors'' ($p_i=1$ for $|E_i-E_0|\leq \Delta$ and vanishing otherwise), Gaussian ones ($p_i = e^{-(E_i-E_0)^2/2\Delta^2}$), or more refined cosine filters \cite{sirui2021algorithms, yilun2022classical}. These choices have an important impact on the results, as we shall see.
The window $W$ allows us to introduce the \emph{microcanonical projection}
\begin{equation}
    \label{wawproj}
    \tilde O = W^{\frac 12}O W^\frac 12
 \end{equation}
\emph{whose eigenvalues and eigenvectors have information that apply to all  $n$}, unlike those of $\rho^{1/2n} O \rho^{1/2n}$.
We will study dynamic correlations
\begin{equation}
    \label{regulaW}
    M^{\mbox{\tiny{W}}}_n(\vec t) = \frac 1{\Tr W} \Tr \( \tilde O(t_1)  \tilde O(t_2) \dots \tilde O(t_n)\) \ . 
\end{equation}

\subsection{Results}
We show that the microcanonical operator $\tilde O$ can indeed be used to compute on-shell correlations functions, namely
\begin{align}
\label{13}
  \frac 1{\Tr W} \Tr & \( \tilde O(t_1)  \tilde O(t_2) \dots \tilde O(t_n)\) \leftrightarrow 
    \Tr \left(\rho^{\frac 1n} O(t_1) \rho^{\frac 1n} O(t_2) \dots \rho^{\frac 1n} O(t_n)\right) 
\end{align}
i.e. $ M_n^{\mbox{\tiny{W}}}(\vec t) \leftrightarrow M_n(\vec t)$,
provided that the window $W$ at energy $E_0=E_\beta$ is chosen appropriately.
Our main findings below are summarized as follows:

\begin{enumerate}
    \item \label{ana} The smoothness properties of window filter $W$ are important. We discuss how the box projector lead to  dynamical effects introduced by the window itself, a familiar fact in signal theory. 
    \item \label{delta} The width $\Delta$ of the window should be \emph{small} enough to be a good projector, but \emph{larger} than the characteristic energy scale $ \omega_{\max}$ so as not to affect the physical correlations. One has
    \begin{equation}
    \label{niente}
        \beta \omega_{\max} \ll \beta \Delta \ll  \sqrt N \ ,
    \end{equation}
 ($N$ is the number of degrees of freedom)    a parametrically large choice of $\Delta$ where the window is well defined. 
    \item When the conditions \ref{ana}. and \ref{delta}. are met, the relation in Eq.\eqref{13}
    is given by
\begin{align}
    M_n^{\mbox{\tiny{W}}}(\vec t) & = \frac{\Tr(W^n)}{\Tr(W)} M_n(\vec t) \ ,
\end{align}
where $\frac{\Tr(W^n)}{\Tr(W)} =e^{-\beta^2 \Delta^2  (1-1/n)} /\sqrt n$ in the case of a Gaussian window. 
    \item Having a microcanonical operator $\tilde O$ leads us to consider the associated \emph{microcanonical operator spectral density} defined as
    \begin{equation}
        \rho(\alpha) \equiv \frac 1{\Tr W} \sum_i \delta(\alpha -a_i) \ ,
    \end{equation}
    where $a_i$ are the eigenvalues of $\tilde O$. Actually, we will focus on $\alpha \rho(\alpha)$ which has a good limit for $\alpha \to 0$ and $\Delta$ as in Eq.\eqref{niente}. We further discuss the generating functions of equal-time moments.
    
\item    On-shell correlations  such as \eqref{regulaW} are related to the usual microcanonical average, e.g.
\begin{equation}
    S^{\mbox{\tiny{W}}}_2(t_1-t_2) = \frac 1{\Tr W} \Tr \( W^2 O(t_1)  O(t_2)\) 
\end{equation}
in the frequency domain by 
\begin{equation}
    S^{\mbox{\tiny{W}}}_2(\omega) = \frac{e^{S(E_0-\omega)}}{e^{S(E_{0}-\omega/2)}} M^{\mbox{\tiny{W}}}_2(\omega)\ ,
\end{equation}
which plays the role of a \emph{microcanonical fluctuation-dissipation theorem} 
\cite{dalessio2016from, khatami2013fluctuation}.
\end{enumerate}

We begin our paper by considering the more familiar case of two-time correlation functions  (Section \ref{sec2}). We then introduce free cumulants and discuss results for $n>2$ (Section \ref{sec:largeq}) and we study the spectral properties (Section \ref{sec:spec}).\\

Our results are tested numerically on a paradigmatic model of many-body quantum chaos: the Ising spin chain with 
 longitudinal and transverse field, described by the Hamiltonian 
\begin{equation}
\label{ising}
H = \sum_{i=1}^L w \sigma_i^x + \sum_{i=1}^L h \sigma_i^z + \sum_{i=1}^{L}J \sigma_i^z \sigma_{i+1}^z \ ,
\end{equation}
where $\sigma_{i}^\alpha$ are Pauli matrices on site $i$ in the direction $\alpha=x, y, z$. 
We measure the energies in units of $J$ and set $w=\sqrt 5/2$, $h=(\sqrt 5+ 5)/8$. With periodic boundary conditions, this system is characterized by translational and inversion symmetry. We consider the subsector at fixed $k=0$ momentum and the even inversion subsector.

\section{The example of two-point functions}
\label{sec2}
Most recent studies of quantum ensembles at equilibrium have focused on one-time expectation values, both in the canonical $M_1= \Tr(\rho )) = S_1$ and in the microcanonical ensemble $M^{\mbox{\tiny{W}}}_1= \Tr(W O)/\Tr W$ \cite{dalessio2016from}.  For example, the issue of how to implement a smooth microcanonical filter $W$ to compute efficiently $M^{\mbox{\tiny{W}}}_1$ has been posed lately in Refs.\cite{sirui2021algorithms, yilun2022classical}.
In this section, we ask this question for two-time correlation functions. These have been studied numerically using for instance microcanonical Lancsos methods Refs.\cite{long2003finite, zotos2006microcanonical, okamoto2018accuracy}. 

We start with the regularized and standard thermal canonical two-times functions respectively:
\begin{align}
    \label{F2B}
    M_{2}^{}(t_1-t_2) &=\text{Tr} \left (\rho^{1/2} O(t_1) \rho^{1/2}O(t_2) \right) \ , \\
    S_{2}^{}(t_1-t_2) &=\text{Tr} \left (\rho^{} \, O(t_1) O(t_2) \right)  \ .
\end{align}
Since these depend only on the time differences $t = t_1-t_2$, we may write everything
in terms of independent time differences $t$, or set $t_2=0$, without loss of generality. The same is true for the $n$-th correlation function in Eqs.\eqref{regulabello}-\eqref{canonical} and in what follows we will put $t_n=0$. The on-shell function $M_{2}^{}(t)$ is
related to the canonical thermal average $S_2(t)$ by a shift in imaginary time $S_2(t) = M_2(t+i\beta/2)$, namely
\begin{equation*}
    \Tr \left(\rho O(t) O(0) \right) = 
    \text{Tr} \left (\rho^{1/2} O(t+ i \beta /2) \rho^{1/2}A \right)\ .
\end{equation*}

The two point functions can be expressed using ETH \eqref{ETHq} for two matrix elements \cite{srednicki1999approach}
\begin{align}
\label{eth2}
\overline{|O_{ii}|^2} & \simeq \overline{O_{ii}}\,^2 \ ; \quad
\\
\overline{|O_{ij}|^2} & = e^{-S(E_{ij}^+)} \; F^{(2)}_{e_{ij}^+}(\omega_{ij}) \quad i\neq j \ ,
\end{align}
where we recall that $\omega_{ij}= E_i-E_j$ and $e_{ij}^+= \frac{E_i+E_j}{2N}$ the energy density and  $F^{(2)}_e(\omega) = |f_O(e, \omega)|^2$ with the standard notations \cite{dalessio2016from}. By simple manipulations, one finds

\begin{equation}
\label{S2}
    S_2(t) = [S_1]^2 + k_2(t) \ ,
\end{equation}

where $S_1 = \mathcal O(e_\beta)$ is the one-point function and $k_2(t)$ is the connected part is the \emph{second (free) cumulant}, which encodes all the information at the second order \footnote{We recall that free cumulants coincide with classical cumulants up to $n=3$ \cite{mingo2017free}, so, in this case, the name is redundant, but useful for later understanding.}.  Thanks to the smoothness properties of ETH \eqref{eth2}, it can be written as the constrained trace over different indices
\begin{equation}
\label{k2t}
    k_2(t) = \Tr' \left ( \rho O(t) O(0) \right ) \equiv \sum_{i\neq j} \frac{e^{-\beta E_i}}Z O_{ij}(t) O_{ij}   \ ,
\end{equation}
where we recall that $\Tr'(\bullet)$ denotes the trace of the arguments with all different indices. Its Fourier transform reads
\begin{equation}
\label{k2w}
    k_2(\omega) = F^{(2)}_{\beta}(\omega)e^{-\beta \omega/2 } \ .
\end{equation}
To isolate the on-shell correlation $F^{(2)}(\omega)$, one should look at the following reduced trace of the thermally projected operator
\begin{align}
\begin{split}
\label{k2bar}
       \overline k_2(t)  \equiv \Tr' \left ( \rho^{1/2} O(t) \rho^{1/2}O(0) \right )
        \equiv & \sum_{i\neq j} \frac{e^{-\beta (E_i+E_j)/2}}Z O_{ij}(t) O_{ij}
    \end{split}
    \\  \to 
      \overline k_2(\omega)  & = F^{(2)}_{\beta}(\omega) \ ,  
\end{align}
which thus identifies the building block of two-point dynamical functions. This expression shows that regularization of the correlation functions in $M_2(t)$ (symmetrization in $\beta$) has the effect of removing the exponential factor $e^{-\beta \omega/2}$ in the Fourier transform of the connected trace $\overline k_2(\omega)$. This is why we referred to it as correlation on the energy shell. 

Summarizing, the dynamic features of two-point functions are encoded in the second cumulants. Their structure is given by the  trace $\Tr'$ of the regularized correlator, reduced to different indices, whose Fourier transform yields ETH on-shell correlations $F^{(2)}_{\beta}(\omega)$ [cf.Eq.\eqref{k2bar}].


As mentioned above, ETH correlation functions are  expected to decay fast at large frequencies $\omega \gg 1$. 
Based on the growth of operators for chaotic systems,   a universal exponential decay has been conjectured \cite{parker2019universal} \footnote{
In the case of interacting integrable systems, there is evidence that $F^{(2)}(\omega)$ has instead a Gaussian decay \cite{parker2019universal}.
 }:
\begin{equation}
\label{decay2}
    F^{(2)}_{\beta}(\omega)\sim e^{-|\omega|/\omega^{(2)}_{\max}} \ .
\end{equation}
 Using that $k_2(0)$ is finite, in Ref.\cite{murthy2019bounds} it was shown that 
 \begin{equation}
     \omega^{(2)}_{\max} \leq \frac 2{\beta\hbar} \ .
 \end{equation}
 Therefore, throughout this paper, we shall assume that
 the frequency $\omega\ll\omega^{(2)}_{\max}$ encompass all the interesting physical properties of a system. \\
 
In summary, ETH tells us that the  physical operators have, on the basis of the Hamiltonian, a `band' structure: the width of the band being $O(1)$ in {\em energy}, while the variations parallel to the diagonal are much
smoother, depending  continuously
on energy {\em density}.

\subsection{Window on operator}

We consider the following microcanonical projection
\begin{equation}
\tilde O = W^{\frac 12}O W^\frac 12
\end{equation} 
where $ W= W_\Delta(H-E_0)$ is the generic window filter defined in Eq.\eqref{W}, well-peaked around energy $E_0$ and with a small width $\Delta/E_0\ll 1$.  
The wish that filtered two-point correlator
\begin{equation}
    \label{F2W}
    M_{2}^{\mbox{\tiny{W}}}(t) = \frac{1}{\Tr W}\text{Tr} \left (W O(t) W O \right) \ ,
\end{equation}
reproduces the thermal regularized one [cf. \eqref{F2B}] and in particular the connected part: $\overline k_2(t)  \equiv \Tr' \left ( \rho^{1/2} O(t) \rho^{1/2}O(0) \right ) $.
Hence,  we look at
\begin{equation}
\label{k2barw}
    \overline k_2^{\mbox{\tiny{W}}}(t) \equiv \frac{1}{\Tr W} \text{Tr}' \left ( W O(t) W O\right ) \ ,
\end{equation}
where the trace is constrained to different indices.
 The size $\Delta$ of the window should obey
 \begin{equation}
     \Delta \gg \omega^{(2)}_{\max} \ ,
 \end{equation}  
 in order not to interfere with the band structure of the operator.

Using standard ETH manipulations \cite{rigol2008thermalization} (substituting sums with integrals and expanding
$S(E+\omega/2) + S(E-\omega/2) - S(E) = S(E) - \frac {\omega^2 }4 S''(E) + \dots$),
one may thus re-write Eq.\eqref{k2barw} to leading order in $N$: 
\begin{align}
	\overline k_2^{\mbox{\tiny{W}}}(t)  & = \frac{1}{\Tr W}
    \sum_{i \neq j} p_i p_j e^{-S(E^+_{ij})} F^{(2)}_{e^+_{ij}}(\omega_{ij})e^{i\omega_{ij} t}
    \\ & \simeq
	\frac{1}{\Tr W} \int dE  e^{S(E)} \; p(E+\omega/2) p(E-\omega/2) 
    \int  d\omega \;e^{i\omega t} 
	F^{(2)}_{E}(\omega) \ .
	\label{eq:QUI}
\end{align}
where $p_i = p(E_i)$ is the window function [cf. Eq.\eqref{Wspec}]. 

The integral over $E$ in Eq.\eqref{eq:QUI} is solved by a saddle point, explicit examples will be discussed below. Since $F_E^{(2)}(\omega)$ is of order one, one can evaluate in on the maximum, leading to  
 \begin{equation}
\label{O2Bup}
    \overline k^{\mbox{\tiny{W}}}_2(\omega) =  F^{(2)}_{E}(\omega)\Big |_{\text{max}} \, P_\Delta(\omega) 
\end{equation}
with  
\begin{equation}
\label{O2B}
P_\Delta(\omega) = 
    \frac{1}{\Tr W} \int dE  e^{S(E)} \; p(E+\omega/2) p(E-\omega/2) \ .
\end{equation}

Eqs.\eqref{O2Bup}-\eqref{O2B} represent the link between thermal on-shell correlations $F^{(2)}_{E}(\omega)$ and the ones obtained from the window and are the main result of this section.

To summarize, we have shown that the microcanonical projected observable $WOW$ yields the same ETH 2-point on-shell correlations but multiplied by a function $ P_\Delta(\omega)$ [cf. Eq.\eqref{O2Bup}], which depends on the different energy windows. We now discuss  if and how one can retrieve dynamical correlations.

\subsection{Not all windows are equal}
\label{secNotequal}


\subsubsection{Box window}
\label{secFlat}
A natural choice for the  window is to consider flat ``box projectors'', namely 
\begin{equation}
    \label{flat}
    p(E) = \theta \left ( \frac \Delta 2  - |E-E_0|\right ) \ ,
\end{equation}
where $\theta(x)$ is a step function.
Flat windows have the following property
\begin{align}
    \Tr \left [ W^n\right ] & =   \Tr \left [ W \right ] =  \int dE \; e^{S(E)} p(E) 
    \nonumber \\ & 
 = e^{S(E_0)} \, \frac 2\beta  \sinh (\beta \Delta /2) \ ,
\end{align}
where integral is solved by saddle point by expanding the entropic factor as $e^{S(E)} = e^{S(E_0) + \beta x}$ around $E_0$ with $\beta=S'(E_0)$ and integrating by $x$.

To see \emph{if} we can retrieve information about two-point dynamical correlators, we need to compute $P_\Delta(\omega)$ in Eq.\eqref{O2B}. A short calculation shows that the box projector constraints both the average energy and frequency inside a box, namely
\begin{align}
    \begin{split}
    \label{pEo}
    p(E+\omega/2)&  \, p(E-\omega/2) = \theta(\Delta-|\omega|)\, \theta \left (\frac{\Delta-|\omega|}2-|E-E_0| \right)\ .
    \end{split}
\end{align}

Hence, we perform the integral over energy $E$ in Eq.\eqref{O2B} solving again by a saddle point.  By expanding the entropic factor as $e^{S(E)} = e^{S(E_0) + \beta x}$ and integrating by $x$, we get
\begin{align}
    \label{PdFlat}
   P_\Delta(\omega) 
   & =  \frac{e^{S(E_0)}}{\Tr W} \int_{-  \frac{\Delta-|\omega|}2} ^{\frac{\Delta-|\omega|}2} e^{\beta x} dx \, \theta(\Delta-|\omega|)
    = \frac{\sinh (\beta (\Delta - |\omega |)/2)}{\sinh (\beta \Delta/ 2)}\theta(\Delta-|\omega|) \ .
\end{align}
The effect of the window does not become negligible for any $\Delta$. This is most striking
in the time domain, where  Eq.(\ref{O2Bup}) becomes a convolution between the two-point correlation and a {\em Lorentzian}. In fact, the Fourier transform of Eq.\eqref{PdFlat} reads
\begin{equation}
  P_\Delta(t) = \frac {4\beta}{\sinh( \beta \Delta /2)}
  \frac {\cosh(\beta \Delta /2) -\cos(\Delta t)}{\beta^2 + 4 t^2} \ .
\end{equation}

Therefore, the box projector has the effect of inducing by itself a slow (Planckian) decay in the dynamical correlator at a time-scale $1/\beta$, see e.g. \cite{reimann2016typical}. This can be traced back to the non-analytic factor $\theta(\Delta-|\omega|)$ in Eq.\eqref{pEo} \footnote{We note that this problem is well-recognized in signal processing, where it is known that nonanalytic windows turn out to have a slowly decaying and oscillating Fourier transform.}. Such behaviour could overcome the intrinsic decay of the correlation functions, which is typically exponentially decaying in time. 
On the other hand, the box projector can still be used for studying the dynamics of correlators with slower decay at long times, such as the long hydrodynamic tails of \emph{diffusive observables} \cite{forster2018hydrodynamic}. \\

\begin{figure}[t]
	\centering
          \includegraphics[width=.45\textwidth]{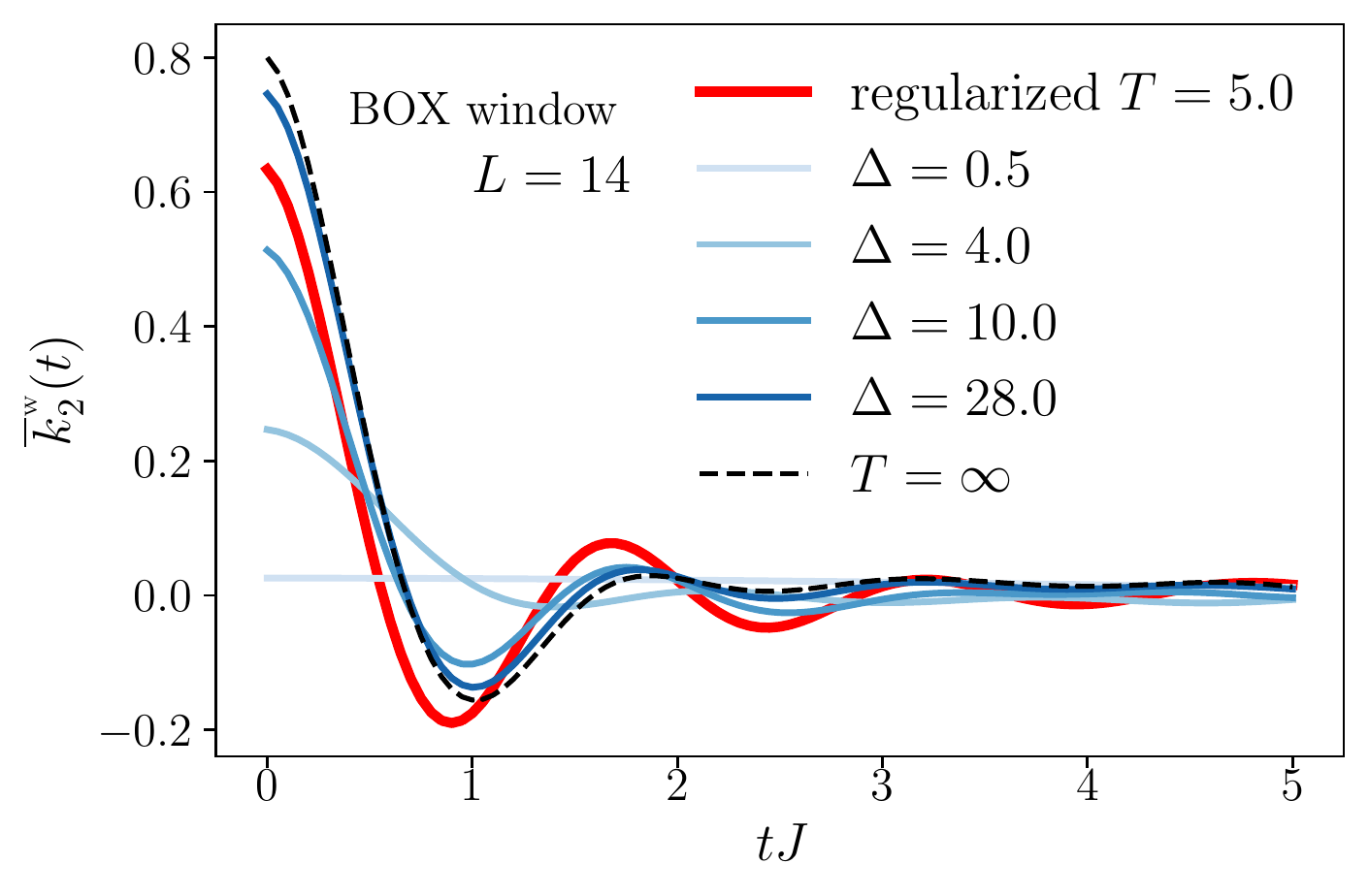}
          \includegraphics[width=.45\textwidth]{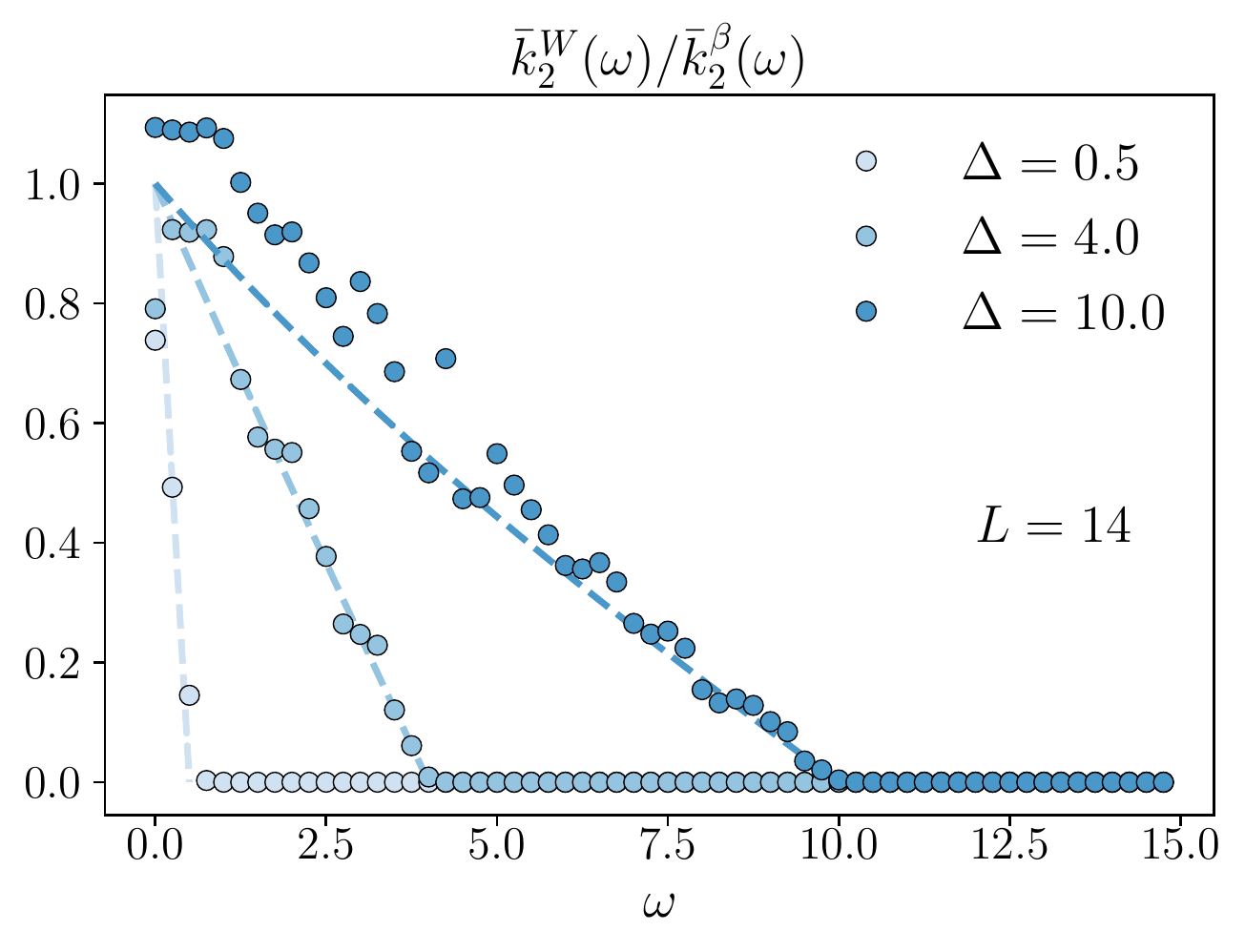}
	\caption{Numerical evaluation of the connected on-shell correlation functions in Eq.\eqref{F2B},\eqref{F2W} with the box projector \eqref{flat} for $O=\frac 1{\sqrt L} \sum_i \sigma_i^z$ with $L=14$.  (Left) The thermally regulated result (red line) is contrasted with one obtained with window projections with different $\Delta=0.5, 4, 10$ (blue lines). When the width is too large $\Delta=2L$ (blue dashed line) one still can not retrieve the result at infinite temperature {which, for very large $\Delta$, should hold independently of where the window is centered} (black line).   (Right) The ratio between the thermal correlation and the regulated thermal ones (dots) compared with Eq.\eqref{PdFlat} without fitting parameters (dashed line). }
	\label{fig:O2t}
\end{figure}

\begin{figure}[t]
	\centering
       \includegraphics[width=.45\textwidth]{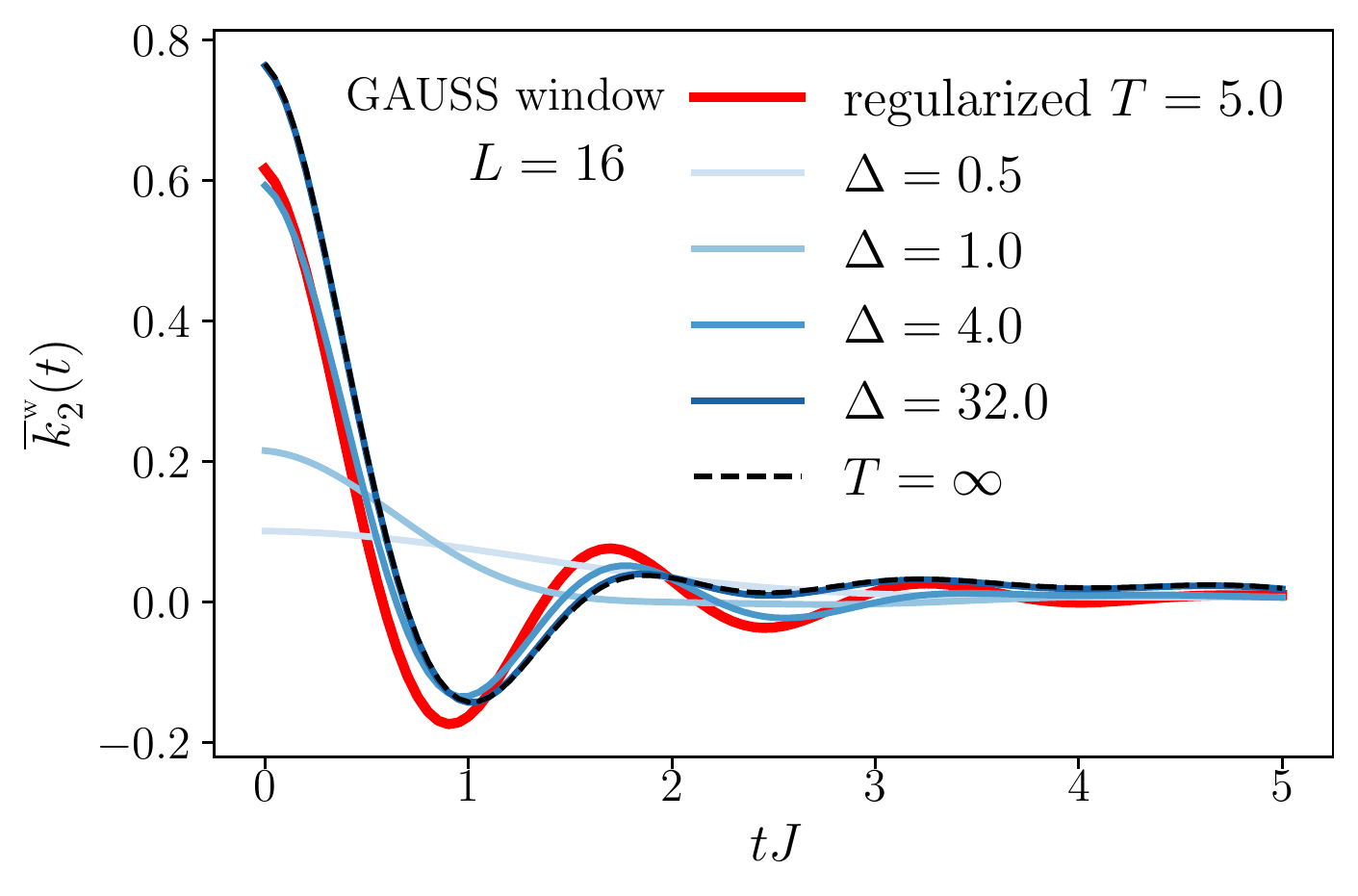}
	\includegraphics[width=.45\textwidth]{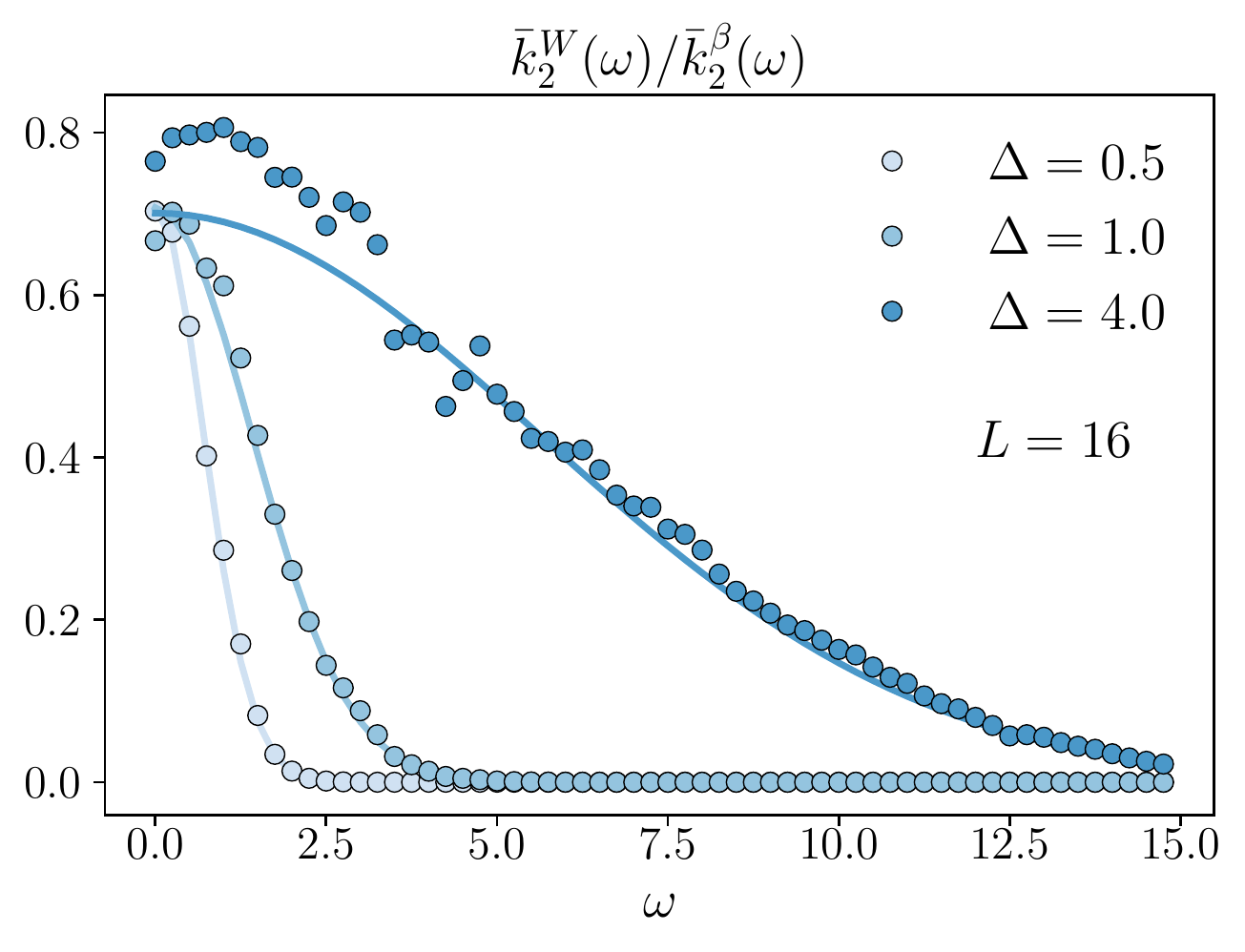}
	\caption{Numerical evaluation of the connected on-shell correlation functions in Eq.\eqref{F2B}-\eqref{F2W} with the gaussian Window \eqref{gaussian} for $A=\frac 1N \sum_i \sigma_i^z$ and $L=16$. (Left) The thermally regulated result (red line) is contrasted with one obtained with window projections with different $\Delta=0.5, 1, 3.5$ (blue lines). When the width is too large $\Delta=2\,L$ (dark blue) one can retrieve the result at infinite temperature (black dashed line).
 (Right) The ratio between the thermal correlation and the regulated thermal ones (dots) compared with Eq.\eqref{pDeltaGAus2} without fitting parameters.  }
	\label{fig:P2w}
\end{figure}

These predictions are consistent with the numerical evaluation for the Ising model, described by the Hamiltonian \eqref{ising}. We focus here on the collective observable $O = \sum\sigma_i^z/\sqrt L$. 
In Fig.\ref{fig:O2t} (left panel), we compare the thermally regulated dynamical correlator (red line) at temperature $T=5$, with the result obtained by WOW obtained with a box window \eqref{F2W}. There is no value of $\Delta$ for which one can reproduce the thermal result. {We also contrast the results with the behavior obtained for infinite temperature,  which should be retrieved for very large $\Delta$ independently of where the window is centered. } Even for very large $\Delta$, one is not able to recover the infinite temperature dynamics (dashed black), which is obtained in the absence of a window filter. {See for comparison Fig.\ref{fig:P2w} below for a smooth window.} This result is induced by the non-analiticities of the window. On the right panel of Fig.\ref{fig:O2t}, 
 we plot the ratio between the Fourier transform of the box energy window and the thermally regulated one for different $\Delta$. The result matches the exact prediction in Eq.\eqref{PdFlat} with no fitting parameter.

\subsubsection{Smooth and Gaussian window}

The long-time tail induced by the box window can be avoided by using a smooth one. As an illustrative example, we consider here Gaussian energy filters:
\begin{equation}
\label{gaussian}
	W = e^{-\frac{(\hat H-E_0)^2}{2\Delta^2}} 
 \ ,\quad \text{i.e.}\quad p(E)= e^{-\frac{( E-E_0)^2}{2\Delta^2}} \ .
\end{equation}
Let us recall, that we want that the entropic term $e^{S(E)}$ to be overridden by $e^{-(E-E_0)^2/\Delta^2}$
 so that the Gaussian dominates in the location of the window. At the same time, the width has to be large enough so as not to spoil the physical correlations of the two-point functions \eqref{decay2}. Hence we need  
 \begin{equation}
     \beta \omega^{(2)}_{\max}\ll \beta \Delta \ll  \sqrt N 
 \end{equation} 
which implies Eq.\eqref{niente}.

To see \emph{how} to retrieve on-shell correlations from the Gaussian window, one needs to evaluate Eq.\eqref{O2B}. First of all, one has
 \begin{equation}
 \label{eq_fac2}
 p(E + \omega/2) p(E-\omega/2)  = p^2(E)\, e^{- \frac{\omega^2\,\,}{2\cdot 2\Delta^2}} \ ,
 \end{equation}
that back in Eq.\eqref{O2B} leads to
\begin{align}
    P_\Delta(\omega) & = e^{- \frac{\omega^2\,\,}{2\cdot 2\Delta^2}} \frac {\Tr [W^2]}{\Tr W} 
     \label{pDeltaGAus2}\ .
 \end{align}  
To simplify this equation, let us compute more generally $  \Tr [W^n]$. By saddle point we have
\begin{align}
    \Tr [W^n] & =  \int dE e^{S(E)} e^{ - n\frac{(E-E_0)^2}{2 \Delta^2}}
     =
    e^{S(E_0)} \int dx e^{- \frac{x^2} {2 (\Delta / \sqrt{n})^2 } + \beta x}
    =  e^{S(E_0)} \frac{\Delta \sqrt {2 \pi}}{\sqrt n} e^{\beta^2 \Delta^2 / n}
    \label{trWq}
\end{align}
where from the left to the right of the first line we have expanded the entropic contribution $e^{S(E)} = e^{S(E_0) + \beta x}$ and then solved the resulting Gaussian integral in $x$. 

All in all, we have shown that, whenever $\Delta$ is large enough [cf. \eqref{niente}], the frequency dependence behaviour of $P_\Delta(\omega)$ can be neglected. We conclude that

\begin{equation}
    \Tr(\rho^{1/2}O(t)\rho^{1/2}O )
    \leftrightarrow
     \Tr( W O(t) W O )
\end{equation}
where the proportionality constant is
$ \frac{\Tr [W^2]}{\Tr[W]} 
    \simeq 
 \frac{ e^{- \beta^2 \Delta^2 / 2}}{\sqrt 2} 
     \label{pDeltaGAus2_}$, which is finite, since $\beta \Delta$ is of order one, even if large.

\section{Higher order correlations}
\label{sec:largeq}
We wish now to access higher order on-shell correlation functions $F^{(n)}_e(\vec \omega)$ defined via Eq.\eqref{ETHq}. In Ref.\cite{foini2019eigenstate}, it was shown that $F^{(n)}_{\beta}(\vec \omega)$ are given by the Fourier Transform of the following constrained trace
\begin{align}
\label{kBeta}
\begin{split}
    &  \overline k_n (\vec t)  = 
    \Tr' \left ( \rho^{\frac 1n} O(t_1) \rho^{\frac 1n} O(t_2) \dots \rho^{\frac 1n} O(0) \right ) 
    \\ & = \frac 1Z \sum_{i_1 \neq \mydots i_n} e^{- \frac \beta q (E_{i_1}+ \dots E_{i_n})} 
    [O(t_1)]_{i_1 i_2}
    \mydots 
    [O(0)]_{i_{n} i_1} 
    = \int d\vec \omega e^{i \vec \omega \cdot \vec t} F^{(n)}_{\beta}(\vec \omega) \ ,   
\end{split}
\end{align}
which corresponds to the connected part of the regularized thermal function in Eq.\eqref{regulabello}, i.e. $ \Tr \left(\rho^{\frac 1n} O(t_1) \rho^{\frac 1n} O(t_2) \dots \rho^{\frac 1n} O(t_n)\right)$. 

These correlations are in fact related to the \emph{thermal free cumulants} \cite{speicher1997free}, a specific form of connected correlation functions defined from the thermal moments as \cite{pappalardi2022eigenstate}
\begin{align}
     S_n(\vec t) & = \text{Tr}(\rho O(t_1) \dots O(t_n)) 
    = \sum_{\pi \in NC(n)} k_\pi(O(t_1) \dots O(t_n)) \ ,   
\end{align}
where $NC(n)$ is the set of the non-crossing partitions of $\{1, 2, \dots, n\}$ and $k_\pi$ is product of cumulants, one for each block of $\pi$.  This decomposition generalizes the implicit definition of connected cumulant $S_2(t) = [k_1]^2+k_2(t)$ [cf. Eq.\eqref{S2}] to higher-order \footnote{For instance 
\begin{align*}
k_1 & \equiv S_1    \\
k_2(t) & \equiv S_2(t) - (S_1)^2\\
k_3(t_1, t_2) & \equiv  S_3(t_1, t_2) - S_1S_2(t_1-t_2) - S_1 S_2(t_1) - S_1 S_2(-t_2)+ 2 (S_1)^3 \\
k_4(t_1, t_2, t_3) & \equiv S_4(t_1, t_2, t_3) - S_2(t_1-t_2) S_2(t_3)  - S_2(t_2-t_3) S_2(t_1)\quad \text{for} \quad S_1=0 \ .
\end{align*}
}. See also Refs.\cite{ebrahimi2016combinatorics, hruza2023coherent} for other appearances of free cumulants in many-body quantum systems.
The smoothness of the ETH ansatz \eqref{ETHq} implies that the thermal free cumulants acquire a simple form, given only by the constrained trace over different indices

\begin{align}
    k_n(\vec t) & = \Tr'(\rho O(t_1) \dots O(t_n)) 
    = \int d\vec \omega e^{i \vec \omega \cdot \vec t} F^{(n)}_{\beta}(\vec \omega) e^{- \beta \vec \omega \cdot \vec \ell_n}  ,
\end{align}

where the right-hand side is the Fourier transform obtained using the smoothness of ETH ansatz. The factor $e^{- \beta \vec \omega \cdot \vec \ell_n}$ is the thermal shift implementing KMS. This expression thus generalizes the $n=2$ case of Eqs.\eqref{k2t}-\eqref{k2w}. To isolate $F^{(n)}_\beta(\vec \omega)$, this result naturally leads us to study the symmetrized correlator $\overline k_n(\vec t)$ in Eq.\eqref{kBeta}, which thus identifies the building block of multi-time functions at all orders $n$. \\


We demand now that  WOW  contains all such correlations. We compute the corresponding object using the window filter:
\begin{widetext}
\begin{align}
    \label{eq:FQ}
	 \overline k_n^{\mbox{\tiny{W}}}(\vec t) 
	 & \equiv 
	 \frac{1}{\Tr W}\hspace{-2.5mm}\sum_{i_1 \neq \mydots \neq i_n} \hspace{-3.5mm}W_{i_1,i_1} \dots W_{i_n,i_n} 
   [O(t_1)]_{i_1 i_2}
    \mydots 
    [O(0)]_{i_{n-1} i_n} 	
\\ & 
= 
	\frac{1}{\Tr W}\int dE e^{S(E)}d\vec \omega \; p(E_1) \dots p(E_n)\; 
	F^{(n)}_{e^+}(\vec \omega ) e^{i \vec \omega \cdot \vec t}  \ .
\end{align}
  \end{widetext}
The same factorization of the probability discussed for $n=2$ [cf. Eq.\eqref{eq_fac2}] applies for generic $n$, namely
\begin{align}
    W_{i_1,i_1}& \dots W_{i_n,i_n} = \; p(E_1) \dots p(E_n)\; 
 = e^{-{ n}(E-E_0)^2/2\Delta^2}
 g_\Delta(\vec \omega) \ ,
 \label{messi}
\end{align}
where  $$ g_\Delta (\vec \omega) = e^{-\frac 1{2\Delta^2} \sum_{i-1}^n \left(\sum_{j=1}^{n-1} c_{j,i} \omega_j \right)^2}\ ,$$  with $c_{j< i} = \frac j n$  \, and $c_{j\geq i}=\frac{j-n}{n}$
depends only on energy differences, as before. 
If we now   take into account that $\beta \Delta \ll \sqrt{N}$, the first factor in (\ref{messi})  dominates over $S(E)$, and
the saddle is $E_0$.

 In order to choose the window appropriately, we recall that, just as the two-point functions [cf.\eqref{decay2}], also $n$-point on-shell correlators are supposed to decay fast at large frequencies, at least as fast as
 \begin{equation}
     F^{(n)}(\vec \omega) = e^{-  |\omega_i|/\omega_{\max}^{(n)}}
 \end{equation}
 in each direction $\omega_i$
with $\omega^{(n)}_{\max} \leq \frac{n-1}{n} \frac 1{\beta \hbar}$ \cite{murthy2019bounds, pappalardi2022eigenstate}. 
From this, we conclude that the largest energy scale characterizing the correlations of the operator $O$  obey
\begin{equation}
\omega_{\max} = \max_n\omega^{(n)}_{\max}\leq \frac 1{\beta \hbar}      \ .
\end{equation}

In the time domain, we thus have
\begin{align}
		\overline k_n^{\mbox{\tiny{W}}}(\vec t)  & = \frac{1}{\Tr W}
		\int dE e^{S(E)} p^n(E)
  \int d\vec \omega \, g_\Delta (\vec \omega) \, 
 \; 
  F_E^{(n)}(\vec \omega)e^{i \vec \omega \cdot \vec t}\\
	& = \frac{\Tr[W^n]}{\Tr[W] }\int d\vec \omega \, g_\Delta (\vec \omega)\, 
  F_E^{(n)}(\vec \omega)e^{i \vec \omega \cdot \vec t} \ ,
\end{align}
where the proportionality constant in the case of the Gaussian can be computed using Eq.\eqref{trWq}, leading to
\begin{align}
\label{Wq}
\frac{\Tr[W^n]}{\Tr[W] }
     = \frac{e^{- \beta^2 \Delta^2 \frac{n-1}n}}{n^{1/2}}   \ .
\end{align}

Thus, by choosing  $\beta \omega_{\max} \ll \Delta \beta \ll \sqrt N$ [cf. Eq.\eqref{niente}], we may neglect the contribution of the filter and set $g_\Delta (\vec \omega) \sim 1$. 
All in all, we have obtained a relation between the time-dependent on-shell correlators and the ones obtained by WOW, i.e. 
\begin{equation}
\label{belloballo}
    \overline k_n^{\mbox{\tiny{W}}}(\vec t) = \frac{\Tr[W^n]}{\Tr[W] } \overline k_n (\vec t) \ .
\end{equation}

Thus, to compute dynamical correlation functions, all the on-shell correlations are encoded in the filtered observable WOW. As a result, the canonical multi-point correlations can be computed directly by looking at their correlations as
\begin{align}
  \label{bellosi}
    \Tr& (\rho^{1/n}O(t_1)\rho^{1/n} \dots O(t_{n-1}) \rho^{1/n} O(0)\rho^{1/n} )
    \leftrightarrow
    \frac 1{\Tr W} \Tr( W O(t_1) \dots  W O(t_{n-1})W O(0) )  
\end{align}
where the proportionality constant is given by Eq.\eqref{Wq}.


\section{The WOW spectrum and generating functions}
\label{sec:spec}
\begin{figure}[t]
	\centering
	\includegraphics[width=.45\textwidth]{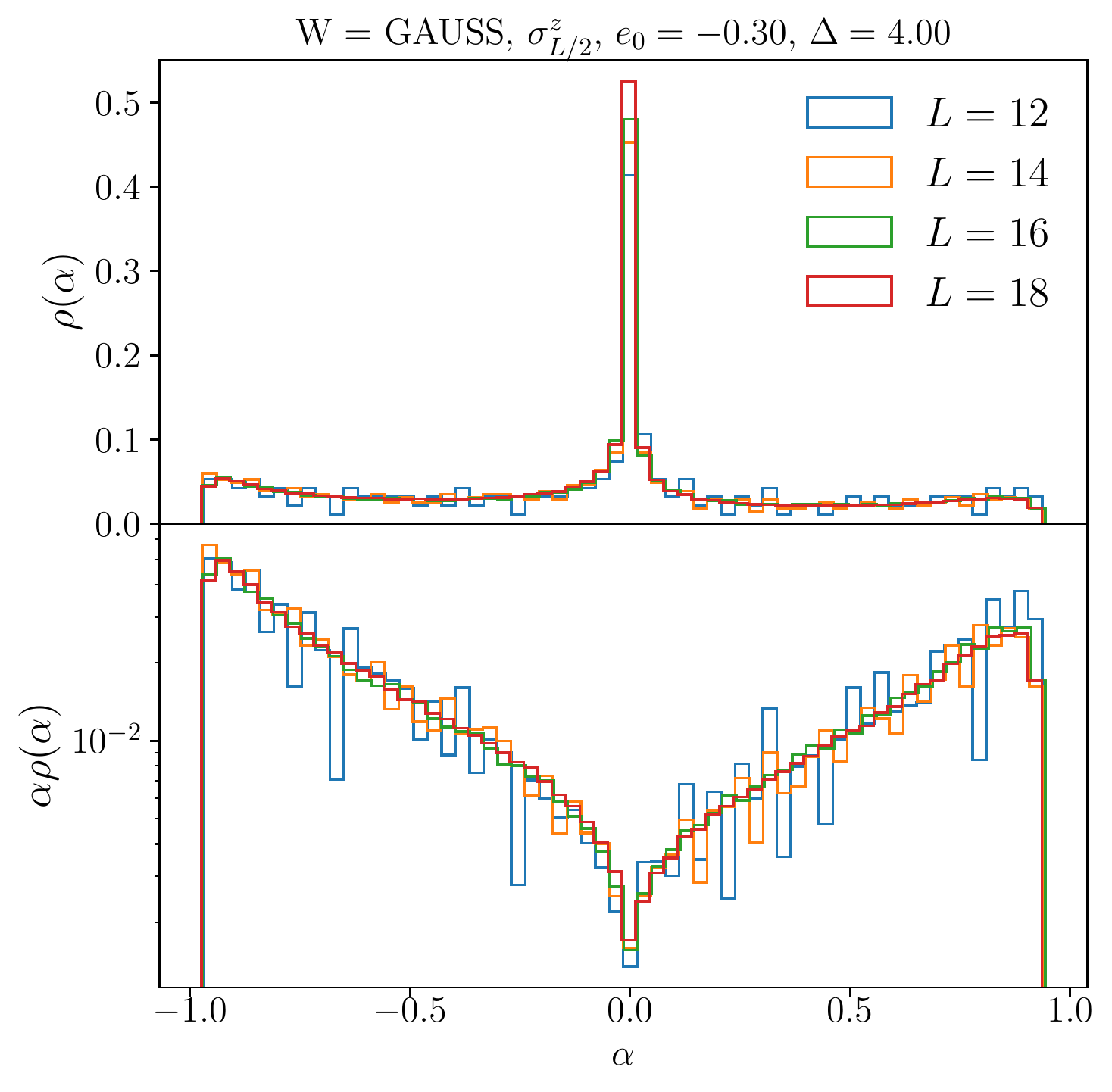}
      \includegraphics[width=.45 \textwidth]{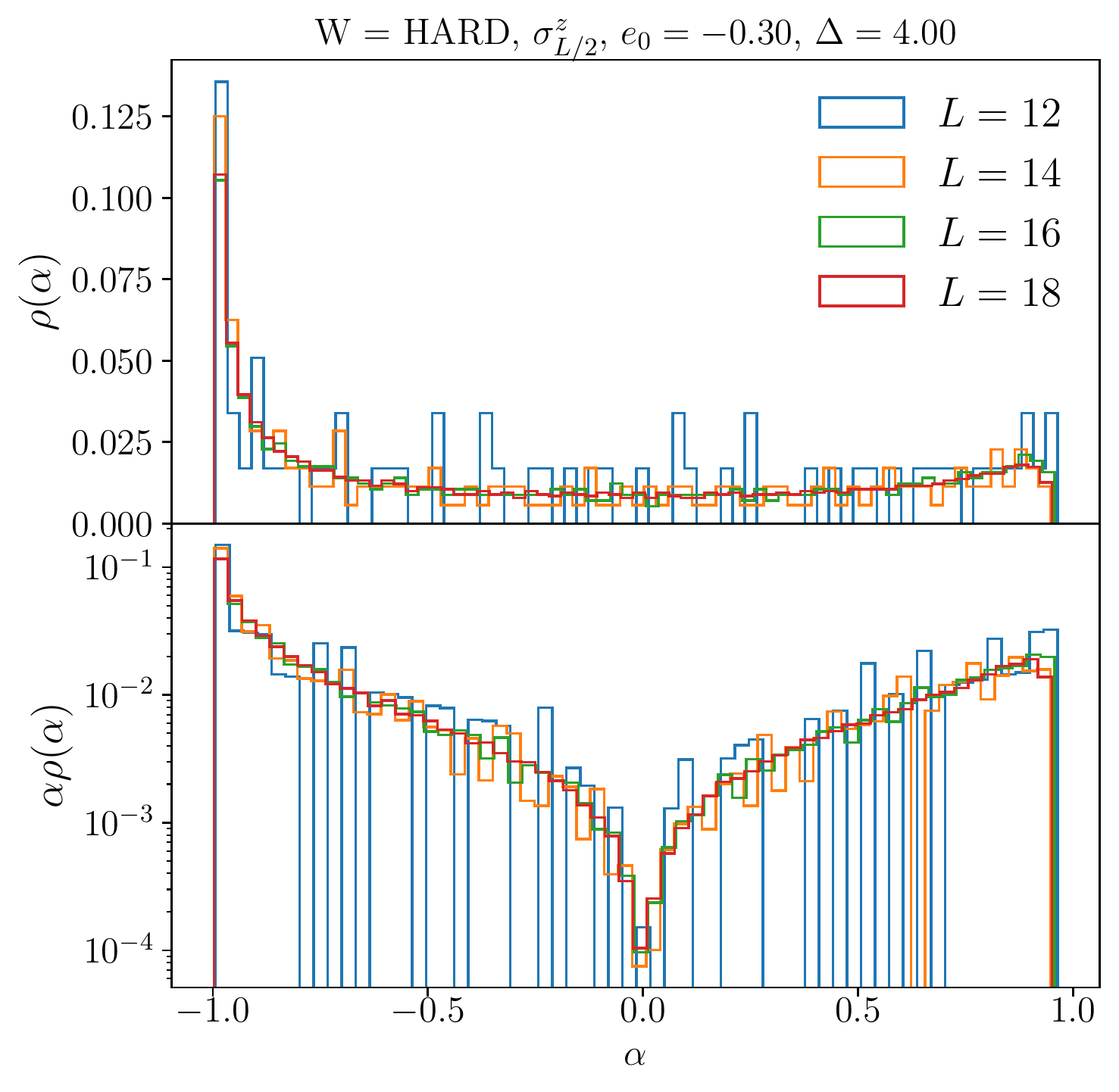}
	\caption{Numerical evaluation of the spectrum of $WOW$ with Gaussian (left) and boxed (right) window filter for different sizes $L=12,14,16,18$. In the bottom panels we plot $\alpha \rho(\alpha)$.}
	\label{fig:spectrum}
\end{figure}

In the previous sections, we have shown that all dynamical on-shell correlations can be retrieved by looking at the structured microcanonical operator WOW. 
We now explore its spectrum.
{The spectrum of the microcanonically truncated operator was introduced and studied by Richter et al. in Ref.\cite{richter2020eigenstate}, to determine the emergence of random matrix description for small windows, i.e. when $\Delta$ is small with the system size $L$. Our approach has a different goal because it aims to access the physical properties of the finite-frequency operator, exactly those that go beyond random matrix theory. }\\
The \emph{WOW spectral density} is defined as
\begin{equation}
\label{rhoa}
    \rho(\alpha) \equiv  \frac 1{\tr W} \sum_{i} \delta(a_i -\alpha) \ ,
\end{equation}
where $a_i$ are the eigenvalues of $\tilde O$. Note that with this definitions $\rho(\alpha)$ is not normalized, i.e. $ \int d\alpha \, \rho(\alpha) 
= \frac{\tr 1 }{\tr W} = e^{S(0)-S(e_0)} $.
In fact, depending on the different choices of the filter W, the distribution $\rho(\alpha)$ could develop a divergent behavior at $\alpha \to 0$. 
Imagine we consider a Gaussian filter as in Eq.\eqref{gaussian}: its effect is to project almost all matrix elements to zero, besides the ones in the energy shall of $E_0$ with variance $\Delta$. Correspondingly, also many eigenvalues will become zero, with the effect of an accumulation in $\rho(\alpha \to 0)$. 
On the other hand, this effect is not present by choosing a box filter function as in Eq.\eqref{flat}. In this case, one can restrain the analysis to the submatrix of $O$ with finite matrix elements, and all the zeros are automatically avoided. {\it Near zero eigenvalues are in any case irrelevant
as far as the integer, positive moments of $\tilde O$ are concerned.}\\
In Fig.\eqref{fig:spectrum}, we compare the numerical histogram of the eigenvalues of $\tilde O$, obtained with a Gaussian filter (left) and the boxed one (right). We consider a  single site observable $O=\sigma^z_{L/2}$, which is characterized by $+1,-1$ eigenvalues in the absence of the microcanonical projection.  As is evident from the plot for the Gaussian filter (Fig.\eqref{fig:spectrum}a), a diverging peak develops in the limit $\alpha \to 0$. This is absent in the case of the box projector (Fig.\eqref{fig:spectrum}c), which displays a regular distribution $\rho(\alpha)$. 


To avoid the pitfalls of the energy filtering for $\alpha \to 0$, we  look at  
\begin{equation}
    \alpha \rho(\alpha) 
\end{equation}
which will appear in the generating function of the finite moments, discussed below. This function of the eigenvalues is a well-defined object, which shall yield a universal result independently of the choice of the window Function $W$. This is shown in Fig.\eqref{fig:spectrum}bd, where we plot $\alpha \rho(\alpha)$ for the Gaussian and the filter functions respectively, which shows a good agreement between the two functions.

\subsection{Generating functions for WOW}

The spectral properties of WOW lead us to the definition of some generating functions to directly access equal-time moments and free cumulants, generalizing known concepts in random matrix theory \cite{bun2017cleaning}.
The equal-time moments are defined as
\begin{equation}
\label{mqdef}
    \overline m_n =  \frac {1}{\Tr[W^n]} \Tr [\tilde O^n] \simeq \Tr \left ( \rho^{1/n} O \dots \rho^{1/n} O \right ) \ ,
\end{equation}
where on the right-hand side we have used the Eq.\eqref{bellosi} derived in the previous section valid for $ \omega_{\max} \beta \ll \Delta \beta \ll \sqrt N$, cf.\eqref{niente}.
Thus, the information of the equal-times moments implies knowledge of the following function 
\begin{equation}
\label{Gbarr}
    \bar G(z) = \sum_{n=0}^\infty \frac{\bar m_n}{z^{n+1}}  = 
    \sum_{n=0}^\infty \frac {1}{\Tr[W^n]}\frac{{ \Tr [\tilde O^n]}}{z^{n+1}} \ ,
\end{equation}
related to the \emph{moment generating function}. 
This expression can always be re-written in terms of the spectral density discussed above, namely
\begin{equation}
\label{Gbar}
     \bar G(z)  \equiv \frac 1z + \frac 1z \int d\alpha \, \alpha \rho(\alpha)  
    \sum_{n=1}^\infty \frac {1}{\Tr[W^n]}\frac{ \alpha^{n-1}}{z^{n}} \ ,
\end{equation}
from which is clear that $\rho$ only appears through $\alpha \rho(\alpha)$ in the expression for the moment $\bar m_n$ generating functions.  \\
In the case of a box window as in Eq.\eqref{flat}, then one has $\Tr[W^n]= \Tr[W]$ and the generating function $ \bar G(z) $ corresponds to the Stieljes (Cauchy) transform  \cite{bun2017cleaning} of the operator WOW:
\begin{align}
       \bar G(z) & = G(z) \equiv \frac 1{\tr W} \tr \left ( \frac 1{z-W^{1/2}OW^{1/2}}\right) 
       = \int d\alpha \frac{\rho(\alpha)}{z-\alpha}  \ .  
\end{align}
Here we use $ G(z)$ to denote the standard definition of the Stieljes transform \cite{bun2017cleaning}, as opposed to $\bar G(z)$ defined in Eq.\eqref{Gbar}. The two correspond only in the case of a box filter, where one can consider a matrix of reduced shape. 
In this case, all the known properties of $G(z)$ can be used. For instance, the spectral density $\rho(\alpha)$ can be retrieve from it by \footnote{Here one uses $\lim_{\epsilon \to 0^+} 1/(x\pm i\epsilon) = \text{p.p.}\frac 1x +\mp i \pi \delta(x)$,  where $p.p$ stands for the Cauchy principal value.}
\begin{equation}
    \rho(\alpha) = \frac 1 \pi \lim_{\epsilon \to 0^+} \text{Im} [G(\alpha - i \epsilon) ] \ .
\end{equation}

\subsection{Other generating functions}
Before concluding, let us add a brief discussion of an alternative  generating function defined from the thermal moments of the observable $O$. 
Standard moments at equal times are defined with respect to the average over the density matrix:
\begin{equation}
    S_n^{\beta}(0) = \text{Tr} (\, \rho_{\beta} \, O^n) \ .
\end{equation}
One can use a standard thermal density matrix $\rho_{\beta}=\frac{1}{Z} e^{-\beta H}$ or a projector over an eigenstate $\rho_{\beta} = |E_\beta\rangle \langle E_\beta |$, the two are equivalent in describing $S_n^\beta$ according to ETH.
Associated with this one can define a Stieltjes transform (resolvent):
\begin{equation}
    G^S_\beta(z) = \text{Tr} \,\left ( \rho_{\beta} \, \frac{1}{z-O} \right )\ .
\end{equation}
Via the Stieltjes one can define an effective probability or density of eigenvalues:
\begin{align}
    \rho^S(x) 
     & = \frac 1 \pi \lim_{\epsilon \to 0^+} \text{Im} [G^S_\beta(x - i \epsilon) ] 
     \nonumber \\ &
     = \frac 1 {\pi}\lim_{\epsilon\to 0} \sum_{\alpha} \langle \alpha | \rho_{\beta} | \alpha \rangle \frac{\epsilon}{(x-O_{\alpha})^2 + \epsilon^2}\ ,
\end{align}
where $|\alpha\rangle$ are the eigenvectors of $O$ and
$O_{\alpha}$ its eigenvalues.
One then recovers that in the infinite temperature limit $\rho \propto I$ this density is simply the density of eigenvalues of $O$, while at finite temperature it gets reweighed with the probability $\langle \alpha | \rho_{\beta} | \alpha \rangle $.
The eigenvalues of an operator can be highly degenerate and in this case, this probability is enhanced by summing over all degenerate eigenvectors.


\section{Discussion}

In this paper, we proposed the construction of a microcanonical observable, which encodes the dynamical correlation functions, and allows us to explore its spectral properties. \\

In order to fix ideas, in this paper we have assumed that the Hamiltonian is sufficiently chaotic, so that the matrix elements of a local operator satisfy the randomicity properties encompassed in the Eigenstate Thermalization Hypothesis. 
It is however very likely that this hypothesis may be relaxed and that one can generalize this construction  to generic microcanonical ensembles which include integrable systems, where one only fixes the energy. 

It would also be interesting to understand how to generalize to Floquet systems.  On one side, one can certainly make a smooth window periodic in the $2\pi$ period Floquet spectrum. 
On the other hand, Floquet systems differ from Hamiltonian ones in that their correlations do not decay exponentially \cite{fritzsch2021eigenstate, pappalardi2023general}, making the $\omega_{\max}$ ill-defined and the filter function definition ambiguous.


\section*{Acknowledgements}
We acknowledge useful discussions with T. Prosen and M. Srednicki, as well as useful suggestions from X. Turkeshi and L. Cugliandolo.
This work was completed during the workshop ``Dynamical Foundation of Many-Body quantum chaos'' held at the Institut Pascal at Université Paris-Saclay made possible with the support of the program “Investissements d’avenir” ANR-11-IDEX-0003-01.
S.P. acknowledges support by the Deutsche Forschungsgemeinschaft (DFG, German Research Foundation) under Germany’s Excellence Strategy - Cluster of Excellence Matter and Light for Quantum Computing (ML4Q) EXC 2004/1 -390534769.

\bibliography{biblio}
\bibliographystyle{unsrtnat}


\end{document}